# Field-Driven Self-Assembly of Magnetite Nanoparticles Investigated Using Small-Angle Neutron Scattering

Zhendong Fu,*[a] Yinguo Xiao,[b] Artem Feoktystov,[a] Vitaliy Pipich,[a] Marie-Sousai Appavou,[a] Yixi Su,[a] Erxi Feng,[a] W. T. Jin[a] and Thomas Brückel[b]

[a.] *Jülich Centre for Neutron Science JCNS, Forschungszentrum Jülich GmbH, Lichtenbergstraße 1, D-85748 Garching, Germany. E-mail: fuzhendong@gmail.com*

[b.] *Jülich Centre for Neutron Science JCNS and Peter Grünberg Institut PGI, JARA-FIT, Forschungszentrum Jülich GmbH, D-52425 Jülich, Germany.*

**Abstract.** The magnetic-field-induced assembly of magnetic nanoparticles (NPs) provides a unique and flexible strategy in the design and the fabrication of functional nanostructures and devices. We have investigated the field-driven self-assembly of core-shell magnetite NPs dispersed in toluene by means of *in situ* small angle neutron scattering (SANS). The form factor of the core-shell NPs was characterized and analyzed using SANS with polarized neutrons. Large-scale aggregation of magnetite NPs has formed above 0.02 T as indicated by very-small angle neutron scattering measurements. Three-dimensional long-range ordered superlattice of magnetite NPs was revealed under the application of moderate magnetic field. The crystal structure of the superlattice has been identified as a face-centred cubic one.

## Introduction

The intriguing phenomenon of self-assembly of colloidal magnetic nanoparticles (MNPs) into well-defined one-, two- and three-dimensional (1D, 2D, 3D) ordered arrays, has been attracting much attention because it provides a bottom-up strategy for the fabrication of functional nanostructures and model systems, which can be manipulated by controlling external parameters such as magnetic or electric field, pressure, temperature, surfactant, and concentration.[1–6] The geometry of the hierarchical structures from self-assembled nanoparticles can also be tailored by controlling the size, shape and interparticle interactions of the constituents.[5,7–10] Ordered arrays of MNPs show different behaviour from that of the bulk and may possess extraordinary application potentials in many fields such as photonics,[11–13] drug delivery and cancer treatment,[14–16] gene transfection,[17,18] patterning,[19–21] energy storage,[22–24] and magnetic levitation.[25,26] As a fast and reversible bottom-up approach among various directed and template-assisted rational strategies, the magnetic field-driven self-assembly of MNPs provides tremendous flexibility and a wide scope for the experimental fabrication.[6,27]

Iron oxide nanoparticles (NPs) are of special interest among the huge number of nanomaterials because of their easy preparation, low cost, high chemical stability, and tunable magnetic and surface properties.[28,29] It is highly desirable to study the self-assembly of iron oxide NPs from both fundamental and application points of view. The 1D chain assembly of $Fe_3O_4$ colloidal nanocrystal clusters was found to show tunable photonic properties across the whole visible region through the application of a relatively weak external field.[11] Dipolar ferromagnetism was revealed in the 2D monolayer of $Fe_3O_4$ NPs with hexagonal packing by Fresnel Lorentz microscopy and electron holography.[30] Large-area 2D assemblies of octahedron-shaped $Fe_3O_4$ NPs were obtained via a simple solvent-evaporation procedure under an in-plane weak magnetic field.[31] A 3D ordering with a base-centred monoclinic symmetry was induced in silica coated hematite nanocubes by an external magnetic field.[32] As to the silica coated $Fe_3O_4$ nanospheres, the equilibrium symmetries of the colloidal crystals were reported to be random hexagonal close packed (RHCP) in absence of external magnetic field and body-centred tetragonal (BCT) with an external field, respectively.[33] Disch *et al.* found that after drop casting the nanoparticle dispersion in an applied magnetic field, both isotropic spherical and anisotropic highly-truncated cubic maghemite NPs are ordered in a face-centred cubic (FCC) arrangement,[34] while a BCT symmetry was obtained in iron oxide nanocubes with moderate degree of truncation.[35]

Small angle neutron scattering (SANS) is one of the most powerful techniques for structural characterization in nanomaterials. Benefiting from its relatively large range of the scattering vector $Q$ ($\equiv 4\pi\sin(\theta)/\lambda$), this technique can provide information not only on the size and shape of the nano-sized constituents by probing their form factor, but also their spatial correlations and organization through the structure factor. Due to the high penetration of neutrons in matter, SANS is well suited for the *in situ* investigations on the samples in liquid.[36,37] The nuclear and magnetic neutron scattering contributions from MNPs can be separated by measuring SANS with polarized neutrons (SANSpol). The very-small angle neutron scattering (VSANS) can detect large aggregations with real-space sizes from several hundred nanometers to several micrometers. In this paper, we employ both techniques to investigate the magnetic-field-driven assembly of $Fe_3O_4$ NPs.

**Experimental**

**Materials**
The suspension of $Fe_3O_4$ NPs in toluene (*i.e.*, ferrofluid) and the toluene for dilution were purchased from Sigma-Aldrich Corporation without any chemical treatment and purification prior to the experiments. The concentration of the as-prepared ferrofluid was 0.6 wt% ($\approx$ 0.1 vol%). The surface of the $Fe_3O_4$ NPs was coated with oleic acid, which allows the dispersity of particles in toluene and prevents the $Fe_3O_4$ NPs from further oxidation.

**Characterizations**
**Transmission electron microscopy (TEM)**. Samples for TEM were prepared by placing a drop of the diluted ferrofluid on a carbon-coated copper grid. After a few seconds, excess solution was removed by blotting with filter paper. Examinations were carried out on a JEM 2200 FS EFTEM instrument (JEOL, Tokyo, Japan) at room temperature with an acceleration voltage of 200 kV. Zero-loss filtered images were recorded digitally by a bottom-mounted 16 bit CCD camera system (FastScan F214, TVIPS, Munich, Germany). Images have been taken with EMenu 4.0 image acquisition program (TVIPS, Munich, Germany) and processed with a free digital imaging processing system ImageJ.[38-40]

**X-ray powder diffraction (XRD).** The XRD measurement was done on a Huber diffractometer with Cu Kα radiation ($\lambda$ = 1.5418 Å) at room temperature. The $Fe_3O_4$ NPs for XRD were obtained by drying the ferrofluid in an argon atmosphere. Data were collected in steps of 0.005° over the $2\theta$ range of 10° – 80°. The background was measured separately and subtracted from the data of the sample.

**Magnetization.** The magnetization of the $Fe_3O_4$ NPs was measured by using a Quantum Design superconducting quantum interference device (SQUID). To prepare the specimen for SQUID, the dried $Fe_3O_4$ NPs were put in a Teflon capsule, which was then inserted in a drinking straw attached to the sample rod of SQUID. The temperature dependence of the magnetization was measured following first a zero-field-cooled (ZFC) and then a field-cooled (FC) protocol. In the ZFC measurement, the sample was cooled from 310 K to 2 K without an external magnetic field, and then the magnetization was measured as a function of temperature under a magnetic field of 500 Oe. Similar procedures were employed in the FC measurement, except that the sample was cooled in a magnetic field of 500 Oe.

**SANS.** The VSANS experiment was carried out on KWS-3[41] instrument running on a double-focusing mirror principle at the Heinz Maier-Leibnitz Zentrum (MLZ) in Garching, Germany. The incident neutron wavelength $\lambda$ was 12.8 Å ($\Delta\lambda/\lambda$ = 20%). The sample-to-detector distance was 5.6 m. The $Q$ range accessible was $6 \times 10^{-4} - 4 \times 10^{-3}$ Å$^{-1}$. The VSANS patterns were recorded by a 2D position sensitive detector with an average pixel size of 0.35 × 0.35 mm. The SANSpol experiment was performed on KWS-1[42] at MLZ in Garching, Germany. The incident wavelength was 5 Å ($\Delta\lambda/\lambda$ = 10%). In order to obtain a higher resolution, we used a collimation length of 20 m and a sample-to-detector distance of 4 m. The SANS patterns were recorded within the $Q$ range between 0.012 – 0.14 Å$^{-1}$. The ferrofluid was put in a quartz cell with dimensions of 30 × 10 × 2 mm in KWS-1 and KWS-3 measurements. The sizes of the sample apertures on both instruments were set as 8 × 8 mm. The SANS and VSANS data presented in this paper are converted to absolute intensity unit of cm$^{-1}$, by means of the data reduction considering the sample thickness, transmission, the scattering from standard samples, and the background from electronic noise, the solvent and the quartz cell.

**Results and discussion**

Fig. 1a shows a typical TEM image and the size distribution of the $Fe_3O_4$ NPs. As revealed by the TEM image, the nanoparticles are spherical in shape and relatively uniform-sized. They show a tendency to form a hexagonal arrangement in monolayer.[43] The average particle diameter was estimated to be 13.7 ± 1.9 nm by counting 300 nanoparticles in several TEM images.

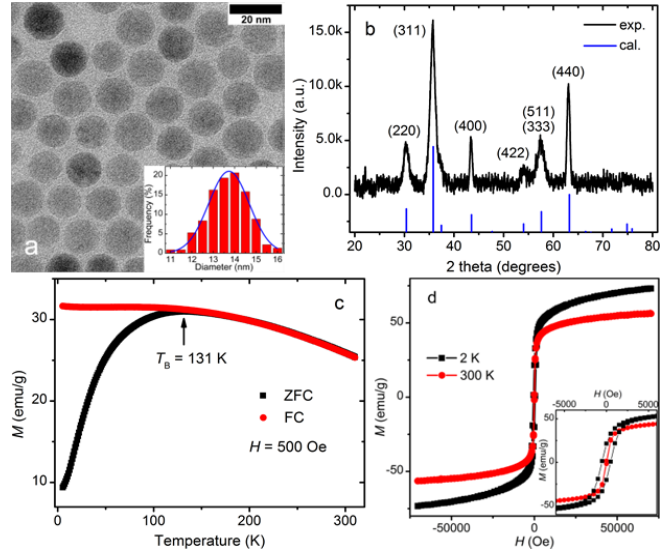

*Fig. 1 (a) Representative TEM image. The inset is the size distribution of $Fe_3O_4$ NPs. (b) Indexed XRD pattern of $Fe_3O_4$ NPs, along with the calculated pattern for a space group of Fd3m (#227) with a lattice constant $a_{Fe3O4}$ = 8.317 Å. (c) Magnetization M vs. temperature plots measured with zero-field-cooled (ZFC, black squares) and field-cooled (FC, red circles) procedures. (d) Magnetization M as a function of the applied magnetic field H measured at 2 K (black squares) and 300 K (red circles). The inset of (d) depicts the zoom-in of the M-H curves at low fields.*

Fig. 1b shows the XRD pattern of the $Fe_3O_4$ NPs, which indicates a cubic spinel structure. The reflection peaks are indexed and agree well with those of $Fe_3O_4$ nanoparticles in the literature.[44] The blue line spectrum in Fig. 1b depicts a simulation of the powder diffraction using the space group *Fd3m* (#227) and a lattice constant $a_{Fe3O4}$ = 8.317 Å, which is close to the lattice parameters reported for magnetite nanoparticles in literature.[45] It can also be seen that the effect of oleic acid on the crystal structure of the core-shell $Fe_3O_4$ NPs is negligible.[46] The average crystallite size of the $Fe_3O_4$ NPs estimated from reflection peaks (220, 311, 400, 511, and 440) by using Sherrer's formula is about 9.4 nm, in agreement with the particle diameter obtained from TEM images.

The temperature dependence of the ZFC and FC magnetization *M* is shown in Fig. 1c. When temperature decreases from 310 to 2 K, the ZFC and FC curves first coincide with each other, and then deviate after the ZFC curve reaches its maximum, indicative of a superparamagnetic behaviour. The so-called blocking temperature $T_B$ is determined to be 131 K, where the ZFC curve exhibits its maximum. At $T_B$, the thermal energy is comparable to the anisotropy energy for flipping the magnetization of the NPs. When the temperature decreases below $T_B$, the magnetization of each NP is blocked in a particular crystallographic direction with the minimum energy. Therefore the ZFC magnetization decreases gradually due to the random orientation of the NPs, while the FC magnetization shows basically a plateau arising from the dipole-dipole interactions between the NPs. When $T > T_B$, the thermal fluctuation of the magnetization of NPs dominates and leads to magnetic behaviours as in a classical paramagnet.[47] As shown in Fig. 1d, the magnetization of $Fe_3O_4$ NPs is plotted as a function of the applied magnetic field (−70000 Oe < $H$ < 70000 Oe) at 2 and 300 K. The absence of magnetic remanence at 300 K is consistent with the superparamagnetic behaviour of $Fe_3O_4$ NPs. The maximum magnetization recorded at 300 K is 56.6 emu/g, which is not saturated even under 70000 Oe and is much smaller than the saturated magnetization of bulk magnetite (92 emu/g).[48] We attribute this phenomenon to the disordered spins on the surfaces of NPs owing to the broken bonds of superficial iron ions and the high surface/volume ratio of the NPs.

Using SANSpol technique, we have investigated both the core-shell microstructure and the field-assisted long-range ordered self-assembly of $Fe_3O_4$ NPs. In a typical SANSpol measurement on KWS-1, the incident neutrons are aligned to be either parallel (-) or antiparallel (+) to the applied field at the sample position. The scattering intensity $I(Q)$ from the sample is the square of the total amplitude and dependent on the polarization state of the incident neutrons. For a dilute system of non-correlated magnetic particles, the scattering intensities as a function of $Q$ are given for two polarization states by[49]

$$I^+(Q,\alpha) = F_N^2 + \{F_M^2 - 2F_NF_M\}\sin^2\alpha, \tag{1}$$

$$I^-(Q,\alpha) = F_N^2 + \{F_M^2 + 2F_NF_M\}\sin^2\alpha, \tag{2}$$

where $\alpha$ is the angle between the scattering vector $\mathbf{Q}$ and the applied magnetic field direction, $F_N(Q)$ and $F_M(Q)$ are the nuclear and magnetic form factors of the magnetic particles, respectively. When $\alpha = 0°$ or $180°$ (*i.e.*, $\mathbf{Q}$ is along the magnetic field), the intensity is independent of the polarization state and originates only from the nuclear contribution. The intensity difference between $I^+(Q,\alpha)$ and $I^-(Q,\alpha)$ represents a cross term of nuclear and magnetic contributions,

$$I^-(Q,\alpha) - I^+(Q,\alpha) = 4F_NF_M\sin^2\alpha, \tag{3}$$

while the average $[I^+(Q,\alpha)+I^-(Q,\alpha)]/2$ corresponds to the scattering of non-polarized neutrons, given by

$$\frac{I^-(Q,\alpha)+I^+(Q,\alpha)}{2} = F_N^2 + F_M^2\sin^2\alpha. \tag{4}$$

The form factor is defined as[50]

$$F(Q) = 4\pi \int_0^{R_{max}} (\eta(r) - \eta_{sol}) \frac{\sin(Qr)}{Qr} r^2 dr \,, \tag{5}$$

where $\eta(r)$ is the scattering length density (SLD) distribution in the particle, $\eta_{sol}$ is the SLD of the solvent, and $R_{max}$ is the outer particle radius. Since our $Fe_3O_4$ NPs are coated with oleic acid, we assume a core-shell model, where the particles contain a core of radius $R$ and a shell of thickness $D$. In this model, the form factor is given by[49]

$$F_{c\text{-}sh}(Q) = [(\eta_c - \eta_{sh})f_{sph}(QR) + (\eta_{sh} - \eta_{sol})f_{sph}(Q(R+D))]V_p \,, \tag{6}$$

where $f_{sph}(x) = 3[\sin(x)-x\cos(x)]/x^3$, $\eta_c$ and $\eta_{sh}$ are the SLD of the core and the shell, respectively.

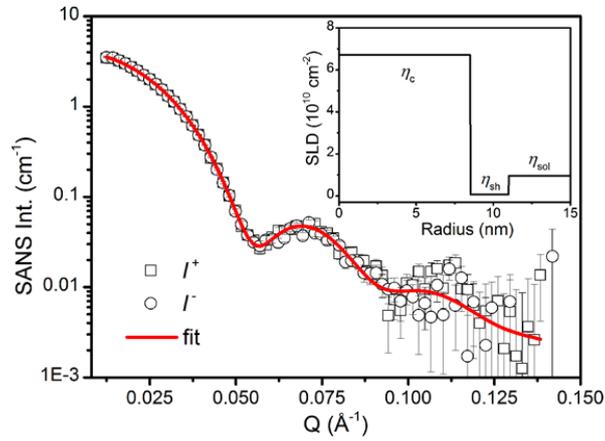

*Fig. 2. SANSpol intensities for diluted $Fe_3O_4$ NP suspension in toluene. The intensities, $I^-$ (circles) and $I^+$ (squares), are measured with flipper-off and flipper-on protocols on KWS-1 at MLZ, respectively. The red line represents the best fit using a core-shell model as described in text. The inset is a schematic drawing of the particle SLD profiles.*

In order to study the microstructure of the $Fe_3O_4$ NPs through the particle form factor, the as-prepared ferrofluid was further diluted to 0.05 vol%. SANSpol measurement was performed on KWS-1 with a small field of 50 Gauss applied at the sample position to keep the neutron polarization. The radially averaged SANS intensity, $I^+(Q)$ and $I^-(Q)$, are plotted in Fig. 2. $I^+(Q)$ and $I^-(Q)$ are nearly coincide with each other, indicating that the magnetic contribution to the total scattering intensity is very small. If the magnetic NPs with a magnetic core and a protonated (H-) surfactant shell are sufficiently diluted in H-solvents, the magnetic scattering contribution to SANS intensity can be neglected.[50] Hence we use the core-shell model given in Eq. (6) to fit directly the $I^+(Q)$ and $I^-(Q)$ curves, which are assumed to contain no structure factors and magnetic contributions. The best fit is shown as the red curve in Fig. 2. The radius of the core and the thickness of the shell determined from the fit are 8.2(5) nm and 2.7(1) nm, respectively. The average particle diameter determined in SANS is thus ~21.8 nm and larger than the one determined in TEM owing to the limited TEM-observable area. The size distribution of $Fe_3O_4$ NPs obtained in the fit is about 15%. In the inset of Fig. 2, a schematic representation is shown for the SLD of the core, the shell and the solvent as a function of radius.

In order to study the formation of $Fe_3O_4$ NP superstructure in magnetic fields, the as-prepared $Fe_3O_4$ ferrofluid was exposed to a vertical magnetic field generated by an electromagnet. The direction of the magnetic field was perpendicular to the incident neutron beam. As shown in Fig. 3, the SANS patterns were collected at various fields ranging from 0.005 T to 2.2 T. Each pixel on the 2D detector of KWS-1 is converted into a vector in reciprocal space with the origin located at the centre of the detector. $Q_x$ and $Q_y$ correspond to the vector components perpendicular and parallel to the magnetic field direction, respectively, while both $Q_x$ and $Q_y$ are normal to the direction of the incident beam. The square-shaped gaps in the SANS patterns are due to the shade of the beam stop. As can be seen in Fig. 3a, the SANS pattern is isotropic at 0.005 T, showing no indication for the presence of locally ordered structures (see also Fig. 4a). When the magnetic field is increased to 0.1 T, clear Bragg peaks appear in Fig. 3b, revealing the formation of single-crystalline-like superstructure. Upon further increasing the magnetic field above 0.1 T (see Fig. 3c – 3f), more particles are aligned due to the stronger dipole-dipole attraction induced by the increasing magnetic field. The crystallinity of the $Fe_3O_4$ NP assembly seems improved as indicated by the clearer high-order diffraction spots, allowing a reliable inspection of the crystal structure.

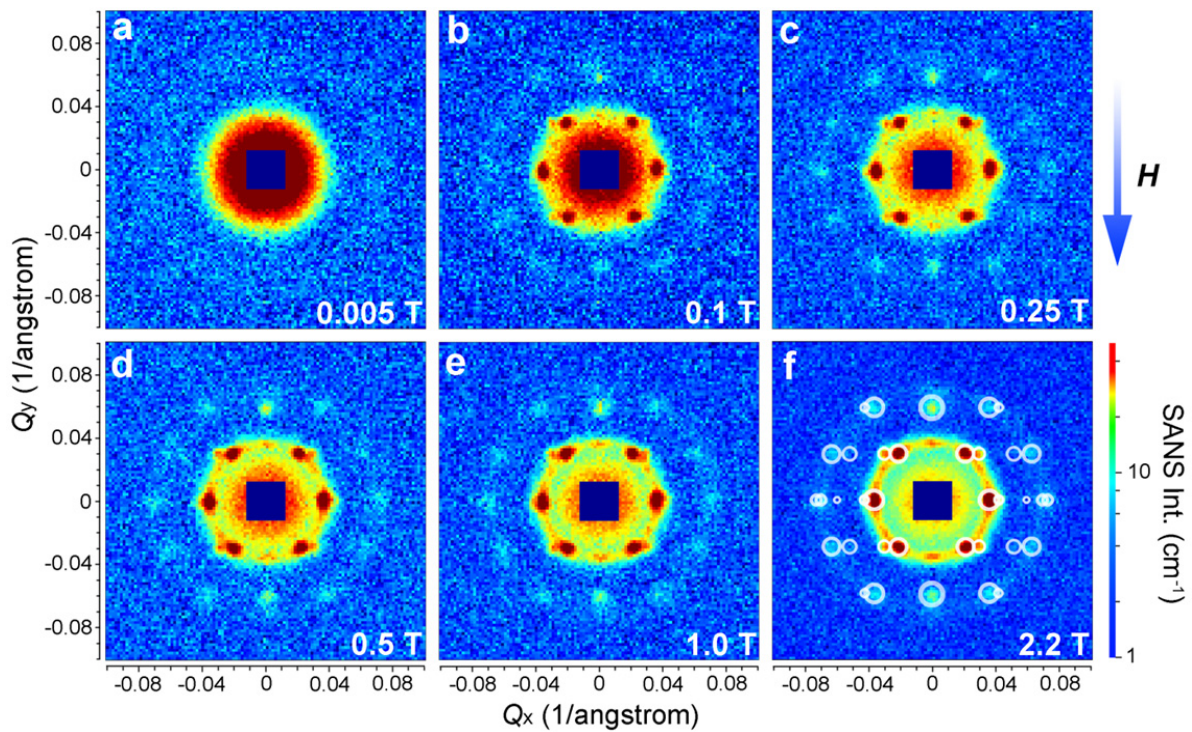

*Fig. 3. SANS patterns of $Fe_3O_4$ ferrofluid exposed to external magnetic fields of 0.005 T (a), 0.1 T (b), 0.25 T (c), 0.5 T (d), 1 T (e), and 2.2 T (f). The vertical magnetic field was aligned perpendicular to the incident neutron beam. The colour bar at the bottom right defines the scale for absolute SANS intensity in unit of $cm^{-1}$. In 3(f), the simulated reflections for face-centred cubic structure are shown as write circles and superimposed over the scattering pattern for comparison.*

When a small magnetic field is applied, the colloidal dipolar particles tend to initially assemble into 1D chains if the dipolar interaction energy is large enough to overcome thermal fluctuations.[3] If the particle concentration and the interparticle magnetic dipole-dipole interaction are further increased, 3D crystalline superstructures of dipolar particles can form. Note that the nearest interparticle distance should be found along the field direction due to the strong dipole-dipole attraction. This means for a certain Bravais lattice the magnetic field defines a special crystallographic direction, along which the nearest neighbors locate. As long as this special crystallographic direction is kept along the field direction, the orientation of crystals is random. As a result, the diffraction intensity is distributed over circles in the reciprocal space, rather than on Brag spots as in the case of a single crystal. Bragg reflections are observed at the intersections of the reciprocal circles and the Ewald sphere surface. In our SANS experiments, we detect only the reflections with the scattering vector $\boldsymbol{Q} = (Q_x, Q_y, 0)$. Since $Q_y$ is in the field direction, the diffraction intensity of each reflection is evenly spread over a circle rotating around the $Q_y$ axis with a radius of $Q_x$. Therefore the measured intensity of an observable reflection should be proportional to the multiplicity $M_{mul}$ of the reflection, but inversely proportional to $Q_x$. The relative intensity of a reflection can then be given by[33]

$$I_{re} \propto \frac{M_{mul}}{Q_x}. \tag{7}$$

In Fig. 4a we plot the radially averaged SANS intensity as a function of $Q$ for various external magnetic fields. It can be seen clearly that long-range ordered self-assembly of $Fe_3O_4$ NPs starts to develop in magnetic fields above 0.1 T. The diffraction peak positions represent a $Q$ ratio of $\sqrt{3} : \sqrt{4} : \sqrt{8} : \sqrt{11}$, corresponding to the (111), (200), (220) and (311) lattice planes of a FCC structure with a lattice constant of $a$ = 29.4 nm. The peak position of (200) planes is extracted from the diffraction pattern in Fig. 3f, because this reflection is hindered by the strong (111) reflection in Fig. 4a. The magnetic field direction defines the [011] crystallographic direction, along which the nearest neighbors in the FCC structure are observed. The distance between the nearest $Fe_3O_4$ NPs can be estimated at 20.8 nm, in good agreement with the particle diameter of 21.8 nm estimated from the diluted sample. The broadening of the diffraction peaks is attributed to the limited crystallite size of the particle assembly. The correlation length is estimated to be about 110 nm by analysing the linewidth (full width at half maximum, FWHM) of the (111) reflection, corresponding to about 5 times the average diameter of the $Fe_3O_4$ NPs. In Fig. 4b, the SANS intensity is integrated in the $Q$ range of 0.0313 – 0.0417 Å$^{-1}$ and plotted as a function of the azimuthal angle between the external field direction and $Q$. The diffraction peaks are enhanced with the increasing magnetic field. The angle distribution of the diffraction peaks quantitatively agrees with that of the (111) reflections of the FCC structure. For example, as shown in Fig. 4b the angle deviation between the two strong peaks in the middle is 71.0°, consistent with the included angle between the $(11\bar{1})$ and $(\bar{1}1\bar{1})$ lattice planes in the FCC structure. Therefore we have simulated the diffraction pattern from a FCC structure and compare it with the measured SANS pattern in Fig. 3f. The simulated data are shown as the white circles. The lattice constant is chosen to be 29.4 nm. As discussed above, an averaging for the crystal orientation has been done in the simulation, while the [011] crystallographic direction is fixed along $Q_y$ by the external field. The radius of the simulated circles is proportional to the square root of the reflection intensity estimated with Eq. (7). An arbitrarily large size has been given to the simulated data with $Q_x = 0$ due to the limited instrumental parameters and crystal disorder.[33] As can be seen in Fig. 3f, the simulated diffraction data agree well with the measured SANS pattern. Note that here we have considered only the structure factor of the $Fe_3O_4$ NP superlattice. Our SANS investigations clearly show that the field-assisted $Fe_3O_4$ NP self-assembly has a FCC type of structure.

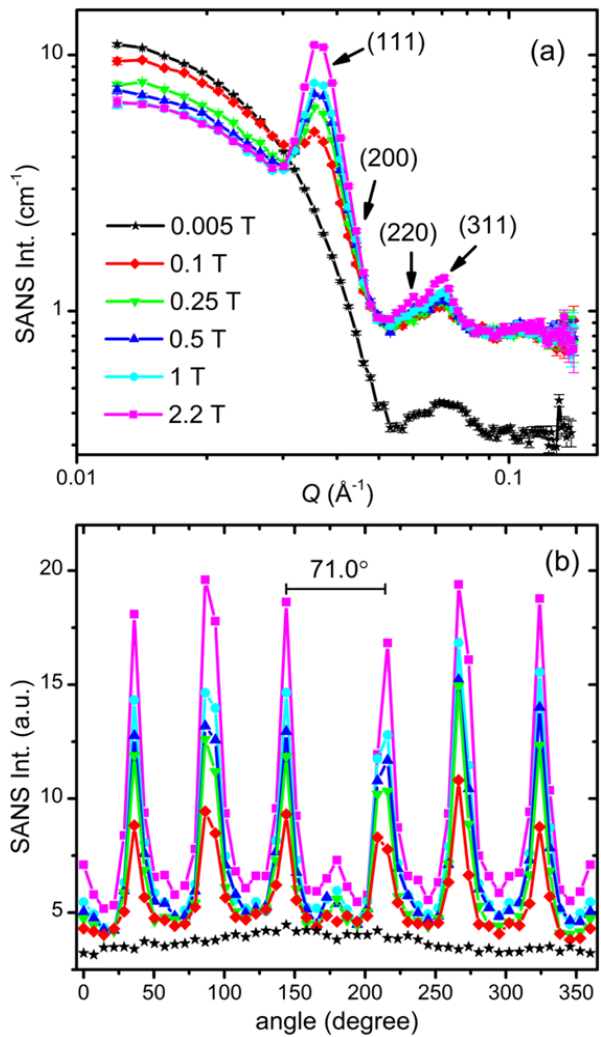

Fig. 4. (a) Radially averaged SANS intensities as a function of Q in various magnetic fields. (b) Field-dependent SANS intensities integrated over $0.0313 \text{ Å}^{-1} < Q < 0.0417 \text{ Å}^{-1}$ as a function of the azimuthal angle between the external field direction and **Q**.

SANSpol measurements were performed on the $Fe_3O_4$ NP self-assemblies at 2.2 T to separate the weak magnetic scattering from the nuclear one. Fig. 5a and 5b show the 2D SANS intensities, $I^-(Q_x, Q_y)$ and $I^+(Q_x, Q_y)$, measured with two neutron polarization states parallel and antiparallel to the external field direction, respectively. Both $I^+$ and $I^-$ show the same clear diffraction peaks as in Fig. 3. In addition to the diffraction peaks, pronounced anisotropy can be seen in the low $Q$ range in both $I^+$ and $I^-$, indicative of the presence of magnetic contribution. Fig. 5c depicts the averaged signal $(I^+ + I^-)/2$, which actually corresponds to the SANS pattern measured with un-polarized neutrons. The difference signal $(I^- - I^+)$ is plotted in Fig. 5d. As explained in Eq. (3), $(I^- - I^+)$ represents the nuclear-magnetic cross term, which shows a clear $\sin^2\alpha$ behaviour with an elongation perpendicular to the field direction. The intensity along the field direction is negligible, because the moments of the $Fe_3O_4$ NPs are aligned parallel to the external field.

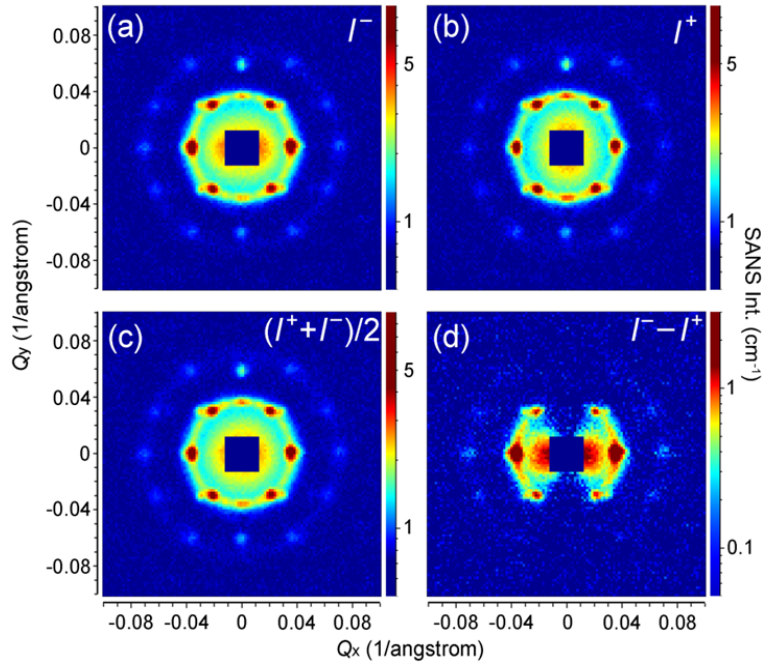

*Fig. 5. SANSpol patterns collected from self-assembled $Fe_3O_4$ NPs in a vertical field of 2.2 T. The neutron polarization direction is either parallel (a) or antiparallel (b) to the external field direction. (c) The averaged signal $(I^+ + I^-)/2$, which corresponds to the SANS pattern measured with un-polarized neutrons. (d) The difference signal $(I^- - I^+)$, reflecting the nuclear-magnetic cross term with a clear $\sin^2\alpha$ behaviour.*

The SANSpol intensities $I^+$ and $I^-$ are integrated over azimuth sectors of 14° in width and plotted as a function of $Q$ in Fig. 6. We choose four azimuth sectors, whose centres are at $\alpha$ = 0°, 35°, 60° and 90°, where $\alpha$ is the angle between $\mathbf{Q}$ and the external field direction. In the 0° sector (Fig. 6a), $I^+$ and $I^-$ are coincide with each other, because the SANS intensity is of nearly pure nuclear origin and thus independent on the neutron polarization. The peak at around $Q$ = 0.036 Å$^{-1}$ exists in both 0° and 60° sectors, and shares the same position with the (111) reflections. This is attributed to the Debye-Scherrer ring corresponding to the (111) reflections from locally misaligned $Fe_3O_4$ NP clusters. The deviation between $I^+$ and $I^-$ increases with increasing $α$, as a direct result of the $\sin^2α$ behaviour of the nuclear-magnetic cross term. Within the resolution of our SANSpol measurements, we do not see a shift of the (111) peaks or other reflections with respect to $α$. Therefore no structure distortion has been detected, although distorted symmetry often occurs in the field-assisted self-assembly of core-shell magnetic NPs.[33,37]

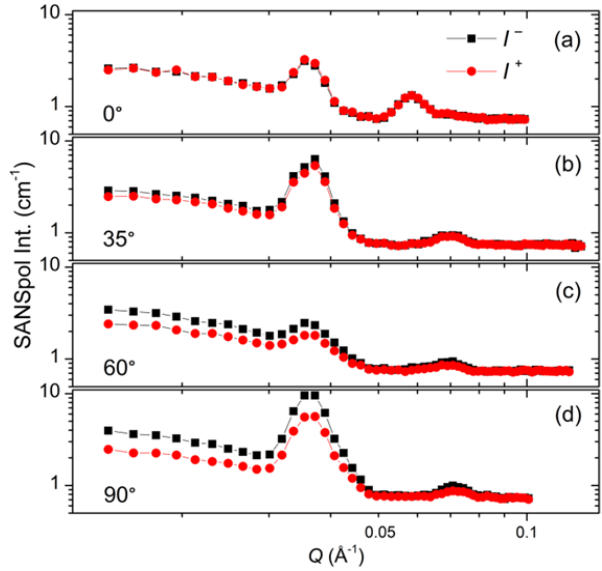

*Fig. 6. SANSpol intensities $I^+$ (red circles) and $I^-$ (black squares) integrated over azimuth sectors of 14° in width. The sector centres are at α = 0° (a), 35° (b), 60° (c) and 90° (d), where α is the angle between the scattering vector **Q** and the applied magnetic field direction.*

The self-assembly process of the as-prepared $Fe_3O_4$ NPs in magnetic fields has been explored by using VSANS with un-polarized neutrons on KWS-3. At the sample position, a vertical field was applied perpendicular to the incident neutron beam. As shown in Fig. 7, the VSANS intensity is integrated over azimuth sectors of 20° in width and plotted as a function of $Q$. The integrated intensity in two sectors centering at 90° and 0° are shown in Fig. (a) and (b) respectively. For the 0° sector, **Q** is along the direction of the external field. Thus the scattering intensity consists of both nuclear and magnetic contributions at low fields, but only nuclear contribution at high fields. However, the SANS intensity nearly shows no field dependence in Fig. 7b, indicating that the magnetic contribution is negligible within the $Q$ range of KWS-3. In Fig. 7a, the integrated intensity over the 90° sector increases as the field increases from 0 T to 0.2 T, and then remains at the same level in the field range of 0.2 – 1.5 T, and finally shows a sharp increase with the field increasing to 2.2 T. For $Q < 0.002$ Å$^{-1}$, the SANS intensities measured in fields above 0.02 T are proportional to $Q^{-4}$, indicative of the presence of large aggregates with smooth surface. Therefore it is confirmed by means of VSANS measurements that large aggregate clusters have already formed at 0.02 T due to the strong dipole-dipole interaction enhanced by the alignment of the dipoles of magnetic NPs in external magnetic field. Such aggregations are much more elongated along the field direction as evidenced by the anisotropy of scattering intensity in 0° and 90° sectors.

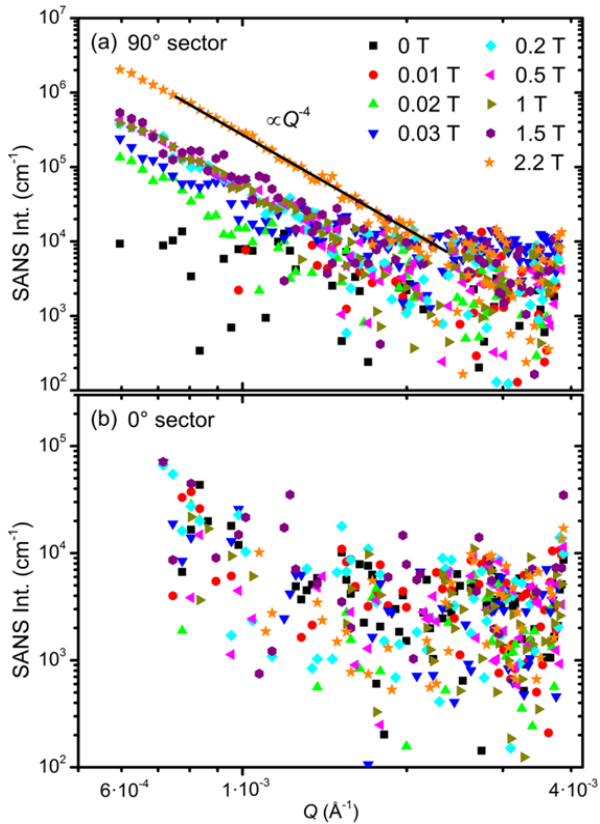

*Fig. 7. Field-dependent VSANS intensities integrated over azimuth sectors of 20° in width. The sector centres are at α = 90° (a) and 0° (b), where α is the angle between the scattering vector **Q** and the applied magnetic field direction. The solid black line in (a) corresponds to a Porod-law behavior, I(Q) ∝ $Q^{-4}$.*

Fundamentally the self-assembly process of magnetic NPs is governed by the interplay between magnetic dipolar interactions, van der Walls forces, Brownian motion and electrostatic repulsive forces. The field-induced strong anisotropy of dipolar forces prefers to align the magnetic dipoles in a head-to-tail configuration, where the dipoles strongly attract each other. However the aligned dipoles are repulsive in a side-by-side configuration. In order to stabilize a 3D close-packed configuration, the aforementioned interparticle interactions must be scaled with each other properly.[51] Although van der Waals attraction in the absence of a magnetic field can already lead to small 2D ordered clusters as suggested in TEM images, a long-range translational order can only be created by an external magnetic field with a sufficient magnitude. Once the magnetic field was applied, the assembly process was completed very fast since we did not see the SANS patterns changing over several hours, consistent with the theoretical predictions.[52] As seen in Fig. 3, the intensity between two (111) reflection peaks cannot be suppressed by enhancing the field. It indicates that there are mosaic-like clusters, which are orientationally misaligned during the initial stage of the assembly process and are blocked later on even in much higher magnetic fields. We attribute these misaligned clusters to the imperfect spherical shapes and the size distribution of the $Fe_3O_4$ NPs.

**Conclusions**

In conclusion, we have investigated the field-driven self-assembly of superparamagnetic core-shell $Fe_3O_4$ NPs dispersed in toluene by means of SANS and VSANS. The form factor of the individual core-shell NP has been measured and analysed. After applying an external magnetic field above 0.2 T, the SANS patterns show that long-range ordered self-assembly of $Fe_3O_4$ NPs is formed. The crystal structure of the NP superlattice has been identified as the face-centred cubic. The VSANS measurements suggest that large aggregations with elongation along the field direction have already appeared at 0.02 T. Our experimental findings shed light on the creation of field-directed self-assembly of colloidal magnetic core-shell nanoparticles into 3D supercrystals, which holds potential for the fabrication of functional nanostructures with novel applications. This work also highlights the superiority of the SANS technique in studying self-assembly phenomena in solution.

**Acknowledgements**

We acknowledge Armin Kriele for the help in usage of the Materials Science Laboratory of Heinz Maier-Leibnitz Centre (MLZ), Technische Universität München (TUM).